\documentclass[final]{aipproc}
\layoutstyle{6x9}
\usepackage{amsmath}
\newcommand{\be}{\begin{eqnarray}}
\newcommand{\ee}{\end{eqnarray}}

\begin{document}

\title{On $\eta$ photoproduction on the neutron and\\ $\pi^0\eta$ photoproduction}



\author{M.~D\"oring}{
  address={Institut f\"ur Kernphysik and J\"ulich Center for Hadron Physics, \\
Forschungszentrum J\"ulich, D-52425 J\"ulich,Germany}
}



\begin{abstract}
The recently discovered enhancement of $\eta$ photoproduction on the quasi-free neutron at energies around 
$\sqrt{s}\sim 1.67$ GeV is addressed within an SU(3) coupled channels model. The $K\Sigma$ threshold plays
a significant role and the quasi-free cross sections on proton and neutron, $\sigma_n$ and $\sigma_p$, 
can be quantitatively explained.
For the reaction $\vec\gamma p\to\pi^0\eta p$, we evaluate the polarization observables $I^S$ and $I^C$ 
from a chiral unitary amplitude developed earlier. 
The $I^S$ and $I^C$ observables have been recently measured for the first time 
by the CBELSA/TAPS collaboration. We show the significance of the 
$\Delta(1700)\,D_{33}$ resonance and its $S$-wave decay into 
$\eta\Delta(1232)$ which further confirms 
the dynamical nature of this resonance. 
\end{abstract}

\maketitle


\section{$\eta$ photoproduction on the neutron}
Recently, the reaction $\gamma n\to \eta n$ has become accessible in photoproduction experiments on the deuteron or nuclei~\cite{Kuznetsov:2007gr,Miyahara:2007zz,Mertens:2008np,Jaegle:2008ux,Kuznetsov:2008hj}. At energies around $\sqrt{s}\sim 1.67$ GeV, an excess of $\eta$ production on the neutron compared to the proton case has been reported in these experiments. 
On the theoretical side, this excess has been interpreted as a potential signal for a non-strange member of an anti-decuplet of pentaquarks~\cite{Diakonov:1997mm,Fix:2007st} although the most prominent of these states, the $\Theta^+(1540)$, may come from a peak created by the experimental cuts, helped 
by a statistical fluctuation due to the limited statistic
of the experiment \cite{MartinezTorres:2010zzb}. For the peak in $\gamma n\to \eta n$, there are also other explanations mostly in terms of different interfering partial waves~\cite{Arndt:2003ga,Choi:2005ki,Shklyar:2006xw,Shyam:2008fr,Chiang:2001as,Anisovich:2008wd}.

In the presented work, the phenomenon is addressed within the chiral unitary framework developed in Ref.~\cite{Doring:2009uc}. Details on the results presented in these proceedings can be found in  Ref.~\cite{Doring:2009qr}, in particular a thorough discussion on the stability of the results. The hadronic interaction in the present framework~\cite{Doring:2009uc,Doring:2009qr} is mediated by the Weinberg-Tomozawa interaction in the coupled channels $\pi N, \,\eta N, K\Lambda$, and $K\Sigma$, unitarized in a Bethe-Salpeter equation. The model also contains explicit resonance states which account for the $N^*(1650)$ and a phenomenological background. The gauge invariant implementation of the photon interaction follows Refs.~\cite{Haberzettl:2006bn,Nakayama:2008tg} (see also Refs. \cite{Borasoy:2007ku}). For the results on the quasi-free $p$ and $n$ in the deuteron, we use the impulse approximation.

In the presented work~\cite{Doring:2009qr}, a global fit of $E_{0+}$ multipoles, $S$-wave cross sections and partial waves has been performed for the reactions $\gamma N\to\pi N$, $\pi N\to\pi N$, $\gamma p\to\eta p$, $\gamma n\to\eta n$, $\pi N\to\eta N$, $\gamma N\to KY$, and $\pi N\to KY$ where $Y=\Lambda,\,\Sigma$, but we concentrate here on the results for $\gamma N\to\eta N$. 
\begin{figure}
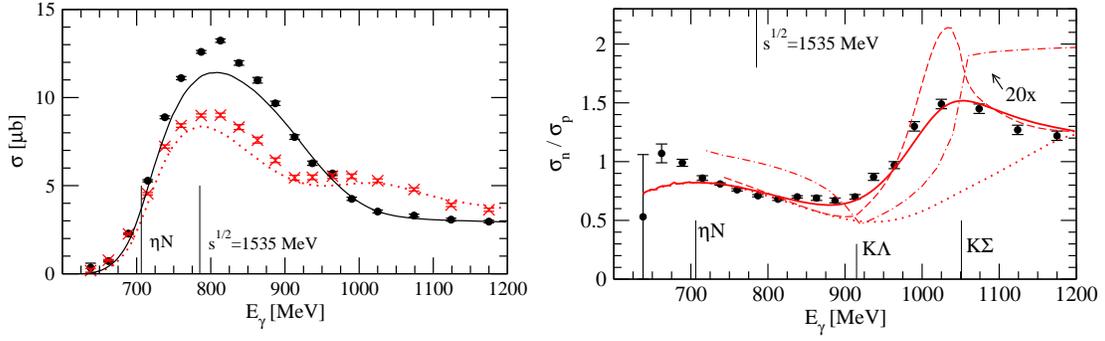

\includegraphics[width=0.47\textwidth]{new_sn_sp_for_paper.eps}
\raisebox{-0.35cm}{
\includegraphics[width=0.5\textwidth]{new_sn_sp_for_paper_ratio.eps}}
\caption{Presented result of Ref.~\cite{Doring:2009qr}. Data: Ref. \cite{Jaegle:2008ux}. 
Left: Cross sections for the photoproduction on the quasi-free proton (solid line) and neutron (dotted line). 
Right: Cross section ratio $\sigma(\gamma n\to\eta n)/\sigma(\gamma p\to\eta p)$. 
Solid (dashed) line: Result, Fermi motion included (excluded). For the other curves, see text.
Also, the $\eta N$, $K\Lambda$, and $K\Sigma$ thresholds are indicated.}
\label{fig:deueta}
\end{figure}
In Fig. \ref{fig:deueta} to the left, the present result is compared with the recent cross section data on the quasi-free $n$ and $p$ from Ref. \cite{Jaegle:2008ux}.  The data are well reproduced. To the right, the ratio of these cross sections is shown (solid line). 
The dashed line indicates the ratio of cross sections on free nucleons, which becomes Fermi smeared (solid line). The appearance of the sharp peak in $\sigma_n/\sigma_p$ is obviously due to the
intermediate states $K\Lambda$ and $K\Sigma$ in the model. Indeed, both in $\gamma n\to\eta n$ and $\gamma p\to\eta p$, the photon can couple to charged pions and kaons in the intermediate $\pi N$ and $K\Sigma$ states; however, in $\gamma p\to\eta p$ the photon coupling to the $K^+$ in the $K^+\Lambda$ state is possible, while this is not possible in $\gamma n\to\eta n$, because the corresponding intermediate state is given by $K^0\Lambda$~\cite{Doring:2009qr}. For the ratio $\sigma_n/\sigma_p$, this difference manifests itself in the observed peak structure in Fig.~\ref{fig:deueta} to the right. Indeed, removing the photon coupling to the $K^+\Lambda$ state in the $\gamma p \to\eta p$ reaction, one obtains the ratio given by the dotted line in Fig. \ref{fig:deueta} to the right; the peak has disappeared.

To check for the model dependence of the presented results, we have replaced the hadronic final state interaction (FSI) with the Weinberg-Tomozawa (WT) term. The photoproduction is then given by the triangle graph which contributes at next-to-leading order (NLO) in the chiral expansion of the amplitude~\cite{Bernard:1994gm}. The resulting ratio $\sigma_n/\sigma_p$, shown as the dash-dotted line in Fig. \ref{fig:deueta} to the right (multiplied by an arbitrary factor of 20), is, of course, very different in magnitude from the full result (dashed line) --- replacing the strong, non-perturbative FSI by the tree-level WT term is certainly an oversimplification. However, the energy dependence shows the same pronounced $KY$ cusps 
as the full result. In the future, the present results could be tested within the hadron exchange framework of Ref.~\cite{Doring:2010ap} where $KY$ states have been included recently.

Summarizing, the experimentally determined $\eta$ photoproduction cross sections on quasi-free neutron and proton can be explained quantitatively within the present model which accounts for the $S$ wave contribution only. The chiral coupled channels SU(3) dynamics and its interplay with the photon lead to the occurrence of the observed spike-like structure in $\sigma_n/\sigma_p$. 


\section{The observables $I^S$ and $I^C$ in the reaction $\vec{\gamma} p \to \pi^0 \eta p$}
The photoproduction of meson pairs is proving to be a rich field allowing us to widen our understanding of hadron dynamics and  hadron structure. 
Following much work devoted to the photoproduction of two pions in the last decade, $\pi^0 \eta$ photoproduction has attracted attention recently~\cite{naka,ajaka,Horn:2008qv,Kashevarov:2009ww}. Polarization observables are reported in Refs.~\cite{ajaka,Gutz:2008zz,Gutz:2009zh,Kashevarov:2010gk}. The reaction was studied theoretically in Refs.~\cite{Jido:2001nt,Doring:2005bx} and 
partial waves analyses of the reaction have been also performed~\cite{Anisovich:2004zz,Fix:2010bd}, finding the $\Delta(1700)$ partial wave to be important.  In the chiral unitary approach of Ref.~\cite{Doring:2005bx},  the process turned out to be dominated at low energies by the excitation of the $\Delta(1700)$, which then decays into  $\eta\Delta $, with the $\Delta$ subsequently decaying into $\pi N$. In that approach, predictions of the cross section are possible because the  $\Delta(1700)\Delta(1232)\eta$ and $\Delta(1700)\Sigma(1385)K$ couplings are known from the chiral unitary framework~\cite{Kolomeitsev:2003kt,Sarkar:2004jh} in which the $\Delta(1700)$ appears dynamically generated. 
In Ref.~\cite{Doring:2007rz} the radiative decay width of the $\Delta(1700) \to\gamma N$ could be predicted, because the photon coupling to the mesons and  baryons that constitute this resonance are all well known. The result is in  agreement with the phenomenologically known values~\cite{Amsler:2008zzb} from  data analyses which are used in Refs.~\cite{ajaka,Doring:2005bx,Doring:2006pt}  for the $\gamma p\to\pi^0\eta p$ reaction.
The theoretical framework from Refs.~\cite{Doring:2005bx,Sarkar:2004jh} is quite predictive since another coupling of the $\Delta(1700)$ resonance is to the  $K \Sigma(1385)$ state and the evaluated (differential) cross sections for the reaction $\gamma p \to K^0 \pi^0 \Sigma^+$~\cite{Doring:2005bx} agree with the measurements published in Ref.~\cite{Nanova:2008kr}. 
In Ref.~\cite{Doring:2006pt}, the chiral unitary amplitude from Ref.~\cite{Doring:2005bx} has been used to relate eleven different pion- and photon-induced reactions. 

\begin{figure}
\raisebox{0.7cm}{
\includegraphics[width=0.3\textwidth]{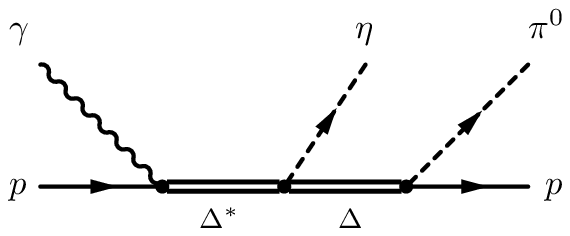}}
 \hspace*{1cm}
\includegraphics[width=0.3\textwidth]{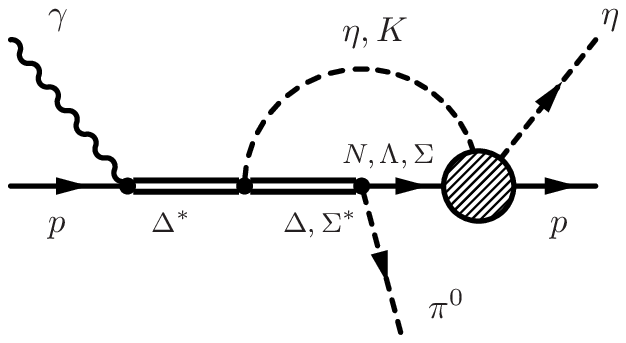}
\caption{Left: Tree level process from the decay of the $\Delta(1700)$ to 
$\eta\Delta(1232)$. This is the dominant process. The complex 
$\Delta(1700)\to\eta\Delta(1232)$ coupling is a prediction within 
the chiral unitary framework. Right: Other processes with $\Delta(1700)\eta\Delta$ and
  $\Delta(1700)K\Sigma(1385)$ 
couplings and $\pi^0\,S_{11}(\eta p)$ final states. For the full list of 
processes, see Ref.~\cite{Doring:2005bx}.}
\label{fig:tree}
\end{figure}

From the list of the underlying processes included in Ref.~\cite{Doring:2005bx}, we show here only those in Fig.~\ref{fig:tree}, which
involve the $\Delta(1700)\eta \Delta$ and  $\Delta(1700)K \Sigma(1385)$ vertices predicted from the chiral unitary  amplitudes~\cite{Sarkar:2004jh}. These processes give the largest
contributions to the $\eta\Delta$ and $\pi^0 S_{11}(\eta p)$ final states.  The latter appears from the unitarization of the meson-baryon amplitude  in which the $N(1535)$ appears
dynamically generated~\cite{Inoue:2001ip}. There are, however, also  processes given by $s$-channel resonance exchange taken from the $\gamma N\to \pi\pi N$ Valencia model of 
Ref.~\cite{Nacher:2000eq}.
Furthermore, there are  contributions from the Kroll-Ruderman and meson pole terms, contributions from the  normal and anomalous magnetic moments of the baryons, and combinations of  those
processes. All free constants that appear in the model have been fixed from other processes, thus the results of Ref.~\cite{Doring:2005bx}  can be regarded as predictions.
The model resulting from Fig.~\ref{fig:tree} is gauge invariant. This is so 
since the $\gamma N \Delta(1700)$ coupling is obtained from the experimental 
data through an expression which is manifestly gauge invariant~\cite{Nacher:2000eq}.

\begin{figure}
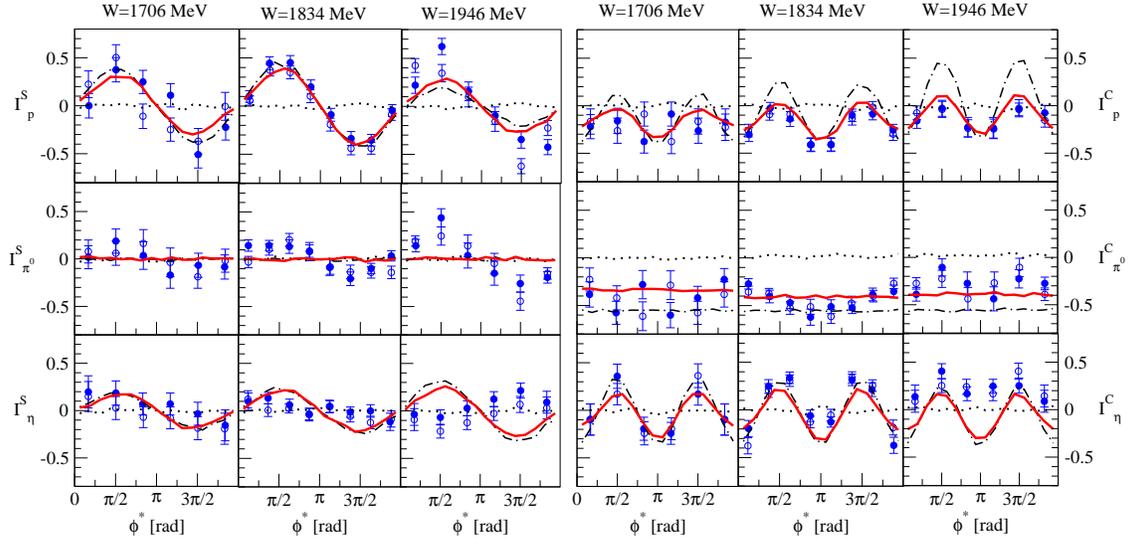

\includegraphics[width=0.5\textwidth]{is_paper.eps}
\includegraphics[width=0.5\textwidth]{ic_paper.eps}
\caption{Results of the presented study~\cite{Doring:2010fw}. Left: Polarization observable $I^S(\phi^*)$, Right:  Polarization observable $I^C(\phi^*)$. Shown are the three cases of $p$, 
$\pi^0$, and $\eta$ spectators and three different energies
$W\equiv\sqrt{s}$. The data are from Ref.~\cite{Gutz:2009zh} 
(full circles). The empty circles show $-I^S(2\pi-\phi^*)$ and $I^C(2\pi-\phi^*)$, respectively. (Red) solid lines:
Present results, predicted from the model of
Refs.~\cite{Doring:2005bx,Doring:2006pt}. (Black) dotted lines: Without the 
processes from 
Fig.~\ref{fig:tree}. (Black) dash-dotted lines: 
Only contribution from the tree level diagram of Fig.~\ref{fig:tree}, left.}
\label{fig:is}
\end{figure}

The recent work of Ref.~\cite{Gutz:2009zh} presents another challenge since new observables are measured, 
i.e, the $I^S$ and $I^C$ polarizations as a  function of the $\phi^*$ angle
between the decay plane and the reaction plane (for the precise definition of the reaction geometry, see e.g. Ref.~\cite{Doring:2010fw}). 
In the work presented here~\cite{Doring:2010fw}, the chiral unitary amplitude developed 
in Refs.~\cite{Doring:2005bx,Doring:2006pt} is used to straightforwardly evaluate $I^S$ and $I^C$ and compare to the data.
The predictions are shown in 
Fig.~\ref{fig:is} with the (red) solid lines, together 
with the data from Ref.~\cite{Gutz:2009zh}.
Note the symmetries $I^S(\phi^*)=-I^S(2\pi-\phi^*)$ and $I^C(\phi^*)=I^C(2\pi-\phi^*)$~\cite{Gutz:2009zh}. 
The present results reproduce well the complex shapes of the angular distributions, while at the highest energies deviations start to become
noticeable. The dotted lines show the results without using the diagrams from Fig.~\ref{fig:tree} which demonstrates that the 
$\Delta(1700)\Delta(1232)\eta$ and $\Delta(1700)\Sigma(1385)K$ couplings provide the essential dynamics. Indeed, using only the 
tree level diagram to the left of Fig.~\ref{fig:tree}, the main features of the full results are already obtained (dashed-dotted line in Fig.~\ref{fig:is}). 

In Ref.~\cite{Doring:2010fw}, also the observables $I^\theta = d\Sigma / d\,\cos\theta$ for the three spectator cases have been evaluated. 
While the agreement with data for the $p$ spectator case~\cite{Gutz:2008zz} is good, there is no published data yet for the other two cases.
In Ref.~\cite{Doring:2010fw}, also a discussion of the theoretical error can be found which turns out to be well under control.
Summarizing, the theoretical predictions of $I^S$, $I^C$, and $I^\theta$ in the reaction $\gamma
p\to\pi^0\eta p$ agree well 
with the data recently measured at CBELSA/TAPS, providing support to a chiral unitary model, in which the $\Delta(1700)$
appears dynamically generated.


\begin{theacknowledgments}
This work is supported by the DFG (Deut\-sche
Forschungsgemeinschaft, Gz: DO 1302/1-2 and SFB/TR-16) and the COSY FFE grant No. 41445282  (COSY-58). 
This work is partly supported by DGICYT and FEDER funds Contract No. FIS 2006-03438, the
Generalitat Valenciana in the program Prometeo and the EU Integrated 
Infrastructure Initiative Hadron Physics Project under contract
RII3-CT-2004-506078. 
\end{theacknowledgments}
\bibliographystyle{aipproc} 

\end{document}